\documentclass[aps,prl,floatfix,twocolumn,superscriptaddress]{revtex4-1}
\usepackage[utf8]{inputenc}
\usepackage{amsmath}
\usepackage{graphicx}
\usepackage{subfig}
\usepackage{color}

\def\<{\langle}
\def\>{\rangle}
\def\(({\left(}
\def\)){\right)}
\def\[[{\left[}
\def\]]{\right]}

\begin{document}

\title{Spin Glass in a Field: a New Zero-Temperature Fixed Point in Finite Dimensions}
\author{Maria Chiara Angelini, Giulio Biroli}
\affiliation{Institut de Physique Th\'{e}orique, CEA/DSM/IPhT-CNRS/URA 2306 CEA-Saclay, F-91191 Gif-sur-Yvette, France}

\begin{abstract}
By using real space renormalisation group (RG) methods we show that spin-glasses in a field display a 
new kind of transition in high dimensions. The corresponding critical properties and the spin-glass phase are 
governed by two non-perturbative zero temperature fixed points of the RG flow. 
We compute the critical exponents, discuss the RG flow and its relevance for three dimensional systems.
The new spin-glass phase we discovered has unusual properties, which are intermediate between the ones
conjectured by droplet and full replica symmetry breaking theories. 
These results provide a new perspective on the long-standing debate about the behaviour of spin-glasses in a field. 
\end{abstract}

\maketitle

%\section{Introduction}
Spin glasses were the focus of an intense and successful research activity in the last forty years. 
The techniques and the concepts developed to understand them had an enormous
impact in several fields. Moreover, spin-glass (SG) theory lead to various spin-offs even in other 
branches of science. Amazingly, despite all these successes 
and forty years of efforts
there is still no consensus on their physical behaviour: the low temperature phase as well 
as the out of equilibrium ageing dynamics remain matter of strong debates.   
On one side, there is the line of research starting from mean-field (MF) models, 
such as the one introduced by Sherrington and Kirkpatrick (SK) \cite{SK}. 
These are solved by the full replica symmetry breaking (FRSB) theory \cite{FRSB},
predicting a SG phase characterized by an infinite number of pure states organised in 
an ultra-metric structure. 
On the other side stands droplet theory (DT), which is a low energy scaling theory 
based on the existence of only two pure states related by spins flip \cite{droplet,dropletBM}.
Although the two approaches provide different predictions, contrasting them 
has proved to be very difficult both in numerical simulations and experiments due
to severe finite-size and finite-time effects \cite{review}. The most clear-cut difference between them concerns
the fate of the SG phase in the presence of an external magnetic field:
the SG phase remains stable up to the so-called de Almeida-Thouless (AT) line $h_{AT}(T)$  
within MF theory \cite{AT}, whereas 
according to the DT it is wiped out by 
even an infinitesimal magnetic field \cite{droplet}. 
In consequence, much of the debate crystallised in proving (or disproving) 
the existence of the AT line in finite dimensional SGs. \\
Field theoretical analysis showed that the Gaussian fixed point (FP) that controls 
the critical behaviour of the AT line for MF model becomes unstable for $d<6$, 
and its basin of attraction shrinks to zero as $d\downarrow6$
\cite{BrayRoberts,BrayMooreSGH} (see also \cite{pimentel}). These findings have two important consequences. First, 
{\it if} there is a transition in a field below six dimension then it necessarily corresponds to a 
non-perturbative (NP) fixed point. Second, this NP-FP could be relevant even well above 
six dimensions: it depends in which basin of attraction the initial condition of the RG 
flow, corresponding to finite dimensional SG, lies. As a matter of fact, the MF behaviour could be recovered 
in very high dimensions only.  
On the numerical side, the most recent numerical results obtained with the use of the
Janus dedicated computer found that no phase 
transition can be identified with traditional data analysis in three dimensions \cite{Janus1,Janus2}; however 
highly non-trivial signals are detected such as a growing correlation
length, peaks in the susceptibility, and a wide probability
distribution function of the overlap, as expected from FRSB. 
Numerical studies performed on one dimensional
long-range (LR) models \cite{LR}, proxies for three dimensional short-range SGs,
support the absence of the AT line, even though there are some particular observables 
that are compatible with a transition in non zero field. 
Finally, RG studies performed by the Migdal-Kadanoff (MK) approximation \cite{MK} also find no SG phase in a field: 
the renormalized couplings initially grow for sufficiently small temperatures and fields 
but eventually vanish when the paramagnetic (PM) FP is reached, as expected from the DT \cite{MooreMK}. 
In conclusion, it is fair to say that the state of the art on SGs in a field, whose study was supposed to 
clarify the situation, is as intricate as the zero-field case \cite{JP-comment}.\\ 
In this work, by using real space RG methods, we show that SGs  
in a field have a new kind of transition for sufficiently large dimensions ($d>8$). 
By studying the RG flow we identify two different zero-temperature FPs, one 
governing the critical properties and the other the low temperature SG phase.
These NP-FPs, whose existence was hinted at by the perturbative RG study discussed above, 
are absent in three dimensions. Nevertheless, they still affect the RG flow and, hence, 
are relevant for the physical behaviour. \\
In the following we present first the analysis performed by the MK-RG method and then 
complement it by using the Dyson hierarchical-RG method \cite{DysonSG}.   
In a nutshell, the MK procedure applied to a hyper-cubic lattice in d-dimensions consists in replacing 
it with a hierarchical diamond one, for which the
MK-RG is exact \cite{MK,Berker}.
Hierarchical diamond lattices (HL) are generated iteratively.
The procedure starts at the step $G=0$ with two spins connected by a single link. 
At each step $G$, for each link of step $G-1$, $p$ parallel branches, made of 2 bonds in series each, 
are added, creating $p$ new spins. 
The relationship between the dimension of the hyper cubic lattice and the number of branches is 
$d=1+\ln(p)/\ln(2)$.
The SG Hamiltonian on HL is the usual one:
$$H=-\frac{1}{p}\((\sum_{\langle i,j \rangle}J_{ij}\sigma_i\sigma_j+\sum_i\sigma_i h_i\)),$$ where $J_{ij}$ and $h_i$ are independent random variables extracted from a Gaussian distribution with variance $v_J^2$ and $v_h^2$ respectively ($v_J=1$ in the following). 
The sum over $i$ and $j$ runs over nearest neighbours on the lattice. Without loss of generality we focus on a random external magnetic field. 
he RG procedure is exactly the opposite of the iterative procedure to construct the HL.
For instance, in step 1, the $p$ spins generated at the last level are integrated out, generating new effective couplings between the remaining spins.  
By integrating out the spins connecting, say $\sigma_1$ and $\sigma_2$, one gets:
 $$\tilde{E}_{1,2}^i=\tilde{J}_{12}^i\sigma_1\sigma_2+ \overrightarrow{h_1}^i\sigma_1+ \overleftarrow{h_2}^i\sigma_2+c_{12}.$$
New fields ($\overrightarrow{h_1}^i$ and $\overleftarrow{h_2}^i$)
associated to each link and an effective coupling between $\sigma_1$ and $\sigma_2$ are generated in addition to a constant $c_{12}$. 
As anticipated, in the presence of external fields there is a difference between HL 
and bond-moving MK. In the MK approximation, the spins in the lattice are divided in blocks of size $\ell$. 
Then all the couplings internal to the blocks are moved to the spins at the edges of the blocks. 
At this point a decimation of the spins at the edges, except those on the corners, is performed.
As for the fields, 
we follow \cite{MooreMK} and move them coherently with the bonds on the spins placed on the edges of the blocks that are traced out in the RG step. In this way, the RG iteration is exactly the same one of a HL except that  
the fields associated with the links are moved from the external spins to the internal ones
for all $p$ branches but one. The unmoved fields represent the ones on the original link.
None of the original site-fields is moved. This change in the renormalisation procedure is important
to have a correct interpretation in terms of bond moving and to avoid pathological behaviours. The exact equations for the flow of the probability distribution of fields and couplings are reported in the Supplementary Information (SI). 
\begin{figure}[t]
\begin{center}
 \includegraphics[width=0.35\textwidth, height=0.18\textheight]{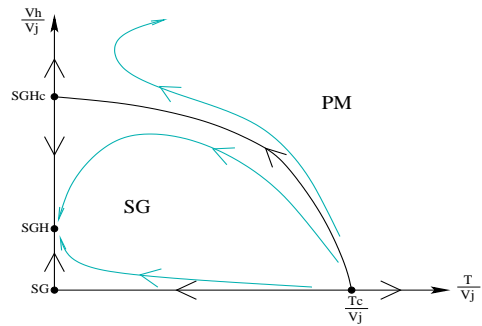}
\caption{Renormalization flow in the plane $T/v_J$-$v_{\vec{h}}/v_J$ for $d\geq d_L$.} 
\label{Fig:RGdiagram}
\end{center}
\end{figure}
We analysed them by using the population dynamics method \cite{PopDyn} (see SI for more details).\\
In the following we present our results on the RG flow. Let us recall first the zero-field results \cite{SouthernYoung}. 
For $d\geq 2.58$ ($p\geq3$), the model has a phase transition from a PM to a SG phase at $T_c^0$. 
The critical temperature is $p$-dependent and is equal to $\frac{1}{\sqrt{p}}$ in the large $p$ limit \cite{gardner}. 
The critical FP related to the transition corresponds to a finite value of $T/v_J$. 
The corresponding SG phase
is associated to a non-trivial zero temperature FP at which the typical value of the couplings after $n$ iteration scale as $v_J^{(n)}\propto\ell^{\theta_0}$, where $\ell$ is the renormalization length after $n$ RG steps: $\ell=2^n$ (see Fig.2). 
The behaviour of $\theta_0$ as a function of the dimension approximatively follows $\frac{d-2.5}{2}$,
which is consistent with the lower critical dimension $d_L=2.5$ found in usual short range SG without field \cite{Boettcher}.
Applying a small field for $T<T_c^0$ the system first approaches the zero temperature FP in zero field but eventually flows away from it since the external field corresponds to a relevant perturbation. Correspondingly, the variances of coupling $v_J$ and bond-field $v_{\overrightarrow{h}}$ 
grow as $v_J^{(n)}\propto\ell^{\theta_0}$, $v_{\overrightarrow{h}}^{(n)}\propto\ell^{\frac{d}{2}}$ with $\theta_0<\frac{d}{2}$, as predicted by the DT. 
The exponent $d/2$ is expected on general grounds because the field couples in a random way to the SG phase. 
No matter how small is the initial value of $v_h$, the renormalised field eventually becomes larger 
than the coupling. On this basis the DT concluded that any infinitesimal field 
destroys the SG phase. This would take place when $v_{\overrightarrow{h}}^{(n)}\propto v_J^{(n)}$ and 
is indeed what we obtain for $d<8.066$.  
In agreement with previous results that focused on the three dimensional case \cite{MooreMK}
one finds that the ratio $v_{\overrightarrow{h}}^{(n)}/v_J^{(n)}$ increases but when it 
exceeds a certain value $r$, 
$v_J^{(n)}$ starts to decrease and $v_{\overrightarrow{h}}^{(n)}$ tends to a constant value.
However for $d\ge d_L=8.066$,
when the ratio ${v_{\overrightarrow{h}}^{(n)}}/{v_J^{(n)}}$ exceeds $r$, 
the growth of $v_{\overrightarrow{h}}^{(n)}$ and ${v_J}^{(n)}$ changes: 
$v_{\overrightarrow{h}}^{(n)}\propto\ell^{\theta}$, $v_J^{(n)}\propto\ell^{\theta}$ with $\theta< \theta_0$.
In Fig. \ref{Fig:RGdiagram} we show the RG flow in the plane $\frac{T}{v_J}$ vs $\frac{v_{\overrightarrow{h}}}{v_J}$. The system flows towards a new zero-temperature stable FP $(T/v_J,v_{\overrightarrow{h}}/v_J)=(0,(v_{\overrightarrow{h}}/v_J)^*)$, called SGH in Fig. \ref{Fig:RGdiagram}, 
which rules the behaviour of the SG phase in a field. 
\begin{figure}[t]
\begin{center}
 \includegraphics[width=0.41\textwidth, height=0.19\textheight]{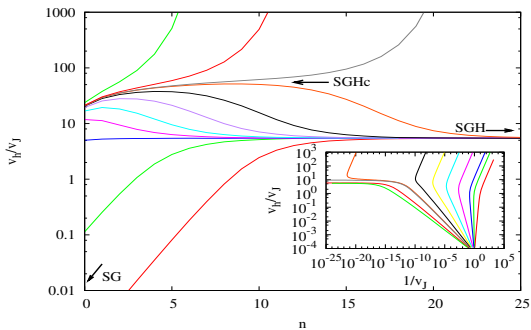}
\caption{Evolution of the observable $v_{\vec{h}}/v_J$ as a function of the renormalization step 
at $T=0$ for $d=10$ starting from different $v_h$. Inset: Renormalization flow in the plane $1/v_J$-$v_{\vec{h}}/v_J$ at $T=0$ starting from $v_h=0.0001$ for dimensions 
$d=2, 2.58, 3, 4, ..., 8, 8.066, 9.23, 9.97$ from right to left.}
\label{Fig:3FP}
\end{center}
\end{figure}
\tiny
\begin{table}[b]
 \begin{center}
 \resizebox{8.5cm}{!} {
  \begin{tabular}{| c | c | c | c | c | c | c | c | c |}
    \hline
    $d$ & $h_c$ & $\theta_{SGH}$ & $\theta_{SGH_c}$ & $\nu$ & $x_{SGH}$ & $x_{SGH_c}$ \\
    \hline \hline
8.066 & 23.4(1) & 0.6222(1) & 0.4833(6) & 9.1(6) & 3.7611(2) & 3.4795(1) \\ \hline
9.229 & 85.15(10) & 1.5824(2) & 0.1203(3) & 1.72(3) & 5.5509(4)  & 2.8640(2) \\ \hline
9.966 & 151.05(10) & 2.0044(9) & 0.060(1) & 1.60(16)& 6.36295(9) & 2.8139(2) \\ \hline 
  \end{tabular}
  }
\end{center}
\caption{Critical values for systems with different $d$.}
  \label{Tab:exponents}
\end{table}
\normalsize

Since at high temperature or for strong fields the system has to flow to 
the PM-FP $(T/v_J,v_{\overrightarrow{h}}/v_J)=(\infty,\infty)$, there is necessarily an unstable FP, that we denote SGH$_c$, separating the disordered (PM) and the ordered (SGH) ones. As shown in Fig. 2, this is also at zero temperature and governs the transition of SGs in a field: when approaching it the couplings and the fields grow as $v_J^{(n)}\propto\ell^{\theta_U}$, 
$v_{\overrightarrow{h}}^{(n)}\propto\ell^{\theta_U}$.
The three zero-temperature FPs are visible in Fig. \ref{Fig:3FP}, where
the evolution of $v_{\overrightarrow{h}}^{(n)}/v_J^{(n)}$ is shown as a function of $n$ at $T=0$, starting
from different initial $v_h$.
We checked that no other FP exists.\\
Now we fully characterise the critical properties. 
For zero-temperature FPs, there are three independent critical exponents, 
one more than for standard phase transitions \cite{BrayMooreRFIM}.
The additional one is $\theta$ that we have already introduced.  
The other two exponents we focus on are $x$ and $\nu$, following the notation of Ref. \cite{BrayMooreRFIM}.
The exponent $x$ describes the rescaling of an infinitesimal symmetry-breaking
field under renormalization, hence it is related to the anomalous dimension of the order parameter. 
The exponent $\nu$ is the one associated to the divergence of the correlation length.
In the case of SGs, the order parameter introduced by Edwards and Anderson \cite{EA} corresponds 
to the overlap between two different replicas subjected to the same quenched disorder. Correspondingly, the symmetry-breaking field $\epsilon$ is an effective attraction (or repulsion) between two different replicas 
$\{{\sigma^1\}}$ and $\{{\sigma^2\}}$. We proceed as for the Random Field Ising Model \cite{machta}:
we introduce a field $\epsilon$ at the extremities of each bond and analyze how its average is renormalized in one RG step: $x=\frac{ln(d\overline{ \epsilon^{R}}/d\epsilon)}{ln(2)}$.
The calculation of $x$ at the zero-field FP can be performed analytically, leading to $x_0=d$.
In order to compute $\nu$ we measured how 
two renormalized flows of the observable $v_{\overrightarrow{h}}/v_J$ corresponding to different original $v_h$
distance themselves.  
The values of $\theta,x$ and $\nu$ as a function of $d$ are reported in Tab. \ref{Tab:exponents};  
all other exponents can be obtained by scaling relations \cite{BrayMooreRFIM}, e.g. $\beta=(d-x)\nu$, $\alpha=2-(d-\theta)\nu$ (we use the standard notation of critical phenomena).  
We find that $\nu$ increases and possibly diverges at $d=8$, as expected since the FPs disappears 
below eight dimension. The fact that $x_{SGH}<d$ implies that the SG phase in a field has a very different nature from its zero field counterpart: the system is ordered but only on a fractal system-size set (accordingly the transition induced by changing $\epsilon$ from $0^+$ to $0^-$ is second order instead of being first order). 
Let us finally discuss the behaviour of correlation functions. 
As it is known for zero temperature FPs, two different
correlation functions are critical \cite{BrayMooreRFIM}.
One is associated to thermal fluctuations:
\begin{equation}
 G_{c}(r)=\overline{\<\sigma_0\sigma_r\>^2-\<\sigma_0\>^2\<\sigma_r\>^2}=\frac{T}{r^{d-2+\eta}}g(r/\xi)
\end{equation}
while the other is associated to disorder fluctuations:
\begin{equation}
 G_{d}(r)=\overline{\<\sigma_0\>^2\<\sigma_r\>^2}-\overline{\<\sigma_0\>^2}\cdot\overline{\<\sigma_r\>^2}=
 \frac{1}{r^{d-4+\tilde{\eta}}}g_{dis}(r/\xi)\,.
\end{equation} 
The exponents $\eta$ and $\tilde{\eta}$ are linked by the relation $\tilde{\eta}-\eta=2-\theta$, and 
$\tilde{\eta}=d+4-2x$. Since $\theta>0$, the 
two correlation functions decay with different power-laws (the disordered one more slowly than the thermal one).
Note that the system is not only critical at the transition, but also in the whole SG phase in a field.\\  
We have also studied analytically the large $d$ limit of the RG equations, as done 
for the zero-field case in Ref. \cite{gardner}. 
We found that $\left.(v_{\overrightarrow{h}}/v_J)^*\right|_{SGH}\simeq5.045$ and 
$\theta_{SGH}(d)\simeq(d-1)/2-2.425$, which are actually good approximations for all dimensions larger than 8.
For $d\rightarrow \infty$, the transition looses its zero temperature character since $\theta_{SGH_c}\rightarrow0$ \cite{footnote}. \\
We now turn to general considerations about our results. First, let us discuss 
their relevance for systems in dimensions less than $d_L$.
In the inset of Fig. \ref{Fig:3FP} we show the flow diagram at $T=0$ for different dimensions, 
starting from a very small field. For $d<d_L$ 
the flow still feels the vestige of the SGH-FP and
is initially attracted towards it, closer and closer as $d$ approaches $d_L$.
However when the ratio $\frac{v_{\overrightarrow{h}}}{v_J}$ becomes larger than the value at the stable FP, 
the transition is avoided and the system finally escapes from the SGH-FP and flows away towards the PM fixed point.\\ 
\begin{figure}[t]
\begin{center}
 \includegraphics[width=0.41\textwidth, height=0.18\textheight]{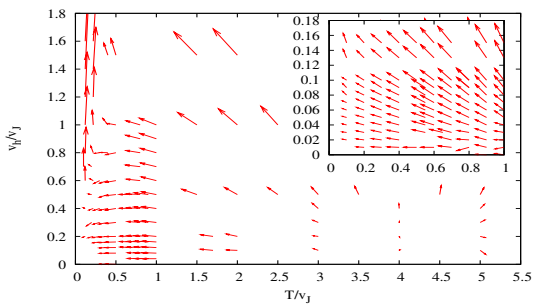}
\caption{Renormalization flow for the Dyson model in dimension $d=20$ ($d=6$ in the inset).}
\label{Fig:Dyson}
\end{center}
\end{figure}
The MK renormalization has pro and cons: for example, it correctly captures the zero temperature FP 
of the Random Field Ising Model \cite{machta}, which is highly non-trivial. On the other hand, it becomes less 
quantitatively accurate in high dimensions and sometimes even fails qualitatively \cite{antenucci}. 
Thus, in order to test the robustness of our results, it is crucial to complement the previous analysis with 
another one that uses a completely different real space RG scheme.
We focus on the one based on  Dyson hierarchical lattice, that is able to 
emulate a short-range model in different dimensions just by changing a parameter. 
We solved the RG equations via an approximation, the real space Ensemble Renormalization Group (ERG) method,
that was introduced and tested for SGs without field in Ref. \cite{DysonSG}. 
Within this framework the relation between original and renormalized parameters (the variance of couplings and fields) 
are obtained by imposing the equivalence between the average of some particular observables over an ensemble of 
$2^n$-spins lattices and an ensemble of $2^{n-1}$-spins lattices \cite{DysonSG} (see SI for more details).
The ERG method was shown to be able to
capture the high dimensional behaviour correctly. For instance, it identifies the upper critical dimension
of SGs in zero field \cite{DysonSG}. Moreover,
it does not suffer of the ambiguity in the treatment of
the external magnetic fields. 
The results we found using essentially the same method and observable as in Ref.\cite{DysonSG} (in particular taking $n=4$) agree with the MK's ones. The corresponding renormalization flow is shown in Fig. \ref{Fig:Dyson} for 
effective dimension $d=20$. A stable zero temperature FP is clearly visible and MF critical behaviour is not recovered. 
In the inset of Fig. \ref{Fig:Dyson} the behaviour for $d=6<d_L$ is shown. As for MK renormalisation, the flow feels the vestige of the SGH-FP and
is initially attracted towards it, however
the transition is avoided and the system finally flows away towards the PM fixed point. \\
In summary, these two very different complementary real-space RG methods lead to the same conclusion:
the initial condition corresponding to microscopic SG models in a field does not lie in the basin of attraction of the Gaussian (MF) fixed point except possibly for very high dimensions. Contrary to what was argued by DT, there is no
general argument against the existence of a transition for SG in a field. 
The flaw in the DT argument is that even though the SG-FP is unstable in presence of an external field, 
the system can nevertheless flow toward a new fixed point SGH$_c$. Indeed, we have unveiled here that 
the very same method used as a basis for DT, the MK-RG approach, shows precisely that in high enough dimensions.\\
The peculiarity of the SG transition in a field is the absence of $\mathcal{Z}_2$ symmetry. 
In consequence in the presence of an external field, 
contrary to the zero field case where it was conjectured the existence of just two pure states (related by spin-flip) 
or an infinite number, the only possibility is the latter one \cite{footnote2}. Whether this is related to FRSB physics is nevertheless unclear. Indeed, we do not find any sign of states characterized by extensive free energy differences of the order of one, a hallmark of FRSB. 
Moreover, the MF transition is not governed by a zero-temperature FP.  
However, it might be that our RG methods are too crude to address this issue. 
For this reason and in order to get a better understanding of the new SG phase discovered in this work and obtain more precise quantitative results (e.g. the value of $d_L$), it would be 
very interesting to develop and apply a more refined non-perturbative RG methods such as the Wetterich's one \cite{wetterich,gilles}. Numerical simulations of high dimensional systems would also be instrumental. 
In particular, it is worth performing new simulations for LR models, proxies of short-range models in large
dimensions. Previous works already found a SG transition in non-zero field for these systems \cite{LRKatzgraber,LRleuzzi}. By using those data, we have analysed the transition of LR models 
corresponding to $d=10$ and compared the quality of mean-field finite size scaling (FSS)
to non mean-field one (in the former case the RG flow is governed by the Gaussian FP, 
whereas in the latter by a non-trivial one). We have found that the non-mean field FSS
is at least comparable, if not even better \cite{footnoteLR}. 
In future analysis it would be interesting to check whether this transition is associated to a $T=0$-FP, in particular
whether disorder fluctuations are much stronger than thermal ones.
The same thing could be checked in short-range models in four dimensions where 
a transition can be identified performing a particular FSS analysis \cite{4dSR}.
As for three dimensional SGs in a field, we notice that the results of numerical simulation can indeed 
be interpreted in terms of an avoided transition where time and length-scales are exponentially related \cite{Janus1,LR},
exactly as it would expected
from the RG flow we obtained. Finally, it would also be interesting to identify the consequences 
of the phase transition we found for the problem of the glass transition,
for which an analogy to the Ising SG in a field was already proposed \cite{moore}.

\begin{acknowledgments}
We acknowledge support from the ERC grants NPRGGLASS. 
We thank J.-P. Bouchaud, A. Decelle, C. Newman, J. Rocchi, D. Stein, G. Tarjus, P. Urbani for useful discussions 
and the authors of \cite{LRKatzgraber,LRleuzzi} for sharing their numerical data.                      
\end{acknowledgments}


\begin{thebibliography}{100}
\bibitem{footnote}{Interestingly, in large $d$ limit, in order to find a finite critical value 
of $v_{\overrightarrow{h}}/v_J$ at zero temperature one has to rescale $v_{\overrightarrow{h}}$ by $p$.
Thus for $d\rightarrow \infty$ the SGH-FP and the one in zero field merge.}
\bibitem{footnote2}{Having just one pure state is in contrast with the existence of a transition.  
The existence of a finite number of pure states is not impossible but seems very unlikely (like having a number between two and infinity in zero field).}
\bibitem{footnoteLR} For example the data of \cite{LRleuzzi} which do not show a FSS crossing for $h=0.2$
for the spin-glass correlation length, do it using non mean field FSS. Moreover the results at higher field, 
such as $h=0.8$, display a slightly better crossing using non mean field FSS.
\end{thebibliography}
\end{document}